\documentclass[prb,aps,onecolumn]{revtex4}

\usepackage{times}
\usepackage{graphicx}
\usepackage{float}
\usepackage{latexsym,amsmath,amssymb,bm,euscript}
\usepackage{color}
\usepackage{subfigure}
\usepackage{epstopdf}
\usepackage[colorlinks=true,linkcolor=blue,citecolor=blue]{hyperref}
\usepackage{type1cm}
\fontsize{12pt}{18pt}

\usepackage{appendix}

\begin{document}

\title{Hydrodynamics of Normal Atomic Gases with Spin-orbit Coupling}
\author{Yan-Hua Hou}
\author{Zhenhua Yu
\footnote{Correspondence to huazhenyu2000@gmail.com}}
\affiliation{Institute for Advanced Study, Tsinghua University, Beijing 100084, China}

\begin{abstract}
Successful realization of spin-orbit coupling in atomic gases by the NIST scheme opens the prospect of studying the effects of spin-orbit coupling on many-body physics in an unprecedentedly controllable way.
Here we derive the linearized hydrodynamic equations for the normal atomic gases of the spin-orbit coupling by the NIST scheme with zero detuning. We show that the hydrodynamics of the system crucially depends on the momentum susceptibilities which can be modified by the spin-orbit coupling. We reveal the effects of the spin-orbit coupling on the sound velocities and the dipole mode frequency of the gases by applying our formalism to the ideal Fermi gas. We also discuss the generalization of our results to other situations.
\end{abstract}
\maketitle

%------------------------------------------------------------------------------
The persisting quest to simulate charged particles in solid state systems by neutral atoms \cite{Dalibard, Zhai12, Zhai15}
stimulated the pioneering experimental achievement of realizing ``spin-orbit" coupling in atomic gases through the Raman process by the NIST group \cite{Lin}. In the NIST scheme, a small magnetic field in the $z$ direction was used to open the degeneracy of atomic hyperfine spins, and two Raman lasers aligning in the $x$ direction shed on a gas of Bose atoms gave rise to a coupling bilinear in momentum and (pseudo-) spin of atoms in one direction. After a unitary transformation \cite{Ho}, the resulting single atom Hamiltonian has the form (we take $\hbar=1$ throughout)
\begin{align}
H_0=\frac{(\mathbf k+k_\text{r}\sigma_z \hat x)^2}{2m}-\frac{\delta}2\sigma_z+\frac{\Omega}2\sigma_x,\label{h0}
\end{align}
where $m$ is the atomic mass, $\sigma_i$ are the atomic (pseudo-) spin operators, $k_\text{r}$ is the Raman laser wave vector, $\delta$ is the detuning tunable by changing the frequency difference between the two Raman lasers, and $\Omega$ is the Rabi frequency for the Raman process. Note that the spin-obit coupling in Eq.~(\ref{h0}) is different from the Rashba and the Dresselhaus forms; when the lasers are turned off, i.e., $\Omega=0$, the coupling $k_x\sigma_z$ in the kinetic energy can be eliminated by a gauge transformation \cite{Ho}. Only nonzero $\Omega$ gives rise to nontrivial spin-orbit coupling. Experimentally $k_\text{r}$ is fixed by the wavevector of the Raman lasers and $\Omega$ can be tuned by the laser intensities. Direct diagonalization of $H_0$ yields the single atom dispersions
\begin{align}
\epsilon_{\pm,\mathbf k}=\frac{\mathbf k^2+k_\text{r}^2}{2m}\pm\sqrt{\left(\frac{k_\text{r}k_x}m-\frac\delta 2\right)^2+\left(\frac\Omega 2\right)^2},\label{dispersion}
\end{align}
with $+$ ($-$) standing for the upper (lower) branch. The same scheme was later applied to Fermi gases \cite{Zhang, Zwierlein}.

The ground state of a Bose gas subject to the spin-orbit coupling realized by the NIST scheme can be either the stripe, the magnetic, or the non-magnetic states depending on the spin-orbit couplings and interatomic interactions \cite{Lin, Ho, Li}. Recently the finite temperature phase diagram of the Bose gas was determined experimentally for zero detuning $\delta=0$ \cite{Ji}. A subsequent perturbative calculation reproduced the correct trend of the thermal effects on the phase boundary between the stripe and the magnetic phases \cite{Yuzq}.

The effects of the spin-orbit coupling of the NIST scheme on the dynamics of atomic gases were first experimentally investigated through the collective dipole oscillation of a Bose-Einstein condensate in a harmonic trap \cite{Chendipole}. In the absence of the spin-orbit coupling, the atomic gas has the Galilean invariance, which guarantees the dipole oscillation frequency $\omega_\text{d}$ equal to the harmonic trapping frequency $\omega_0$ \cite{Stringaridipole}. The dispersions (\ref{dispersion}) given rise to by the NIST scheme apparently break down the Galilean invariance, which indicates that $\omega_\text{d}$ can be different from $\omega_0$. Nevertheless, since during the small dipole oscillation the Bose-Einstein condensate accesses mostly the lower branch states around a minimum of  $\epsilon_{-,\mathbf k}$ at momentum $\mathbf k_0$, it is sufficient to approximate the dispersion $\epsilon_{-,\mathbf k}\approx \epsilon_{-,\mathbf k_0}+(\mathbf k-\mathbf k_0)^2/2m^*$, whose form restores the Galilean invariance though the ``effective mass" $m^*$ depends on the spin-orbit coupling. The experimental data for the dipole frequency of the Bose-Einstein condensate with the spin-orbit coupling can be mainly explained by the ``effective mass" approximation $\omega_\text{d}\sim\omega_0\sqrt{m/m^*}$ \cite{Chendipole}.

Up to now, the study of the collective modes and the hydrodynamics of the spin-obit coupled atomic gases mainly focuses on the Bose-Einstein condensates, in which case the existence of a single condensate wave function greatly simplifies the theoretical treatment \cite{Zheng, Zhai2, Martone, Han, Cooper, Zhang13}. However, to describe the dynamics of Bose gases with a substantial normal fraction and Fermi gases in the presence of spin-orbit coupling requires a more general framework. In this work, we consider atomic gases in the hydrodynamic regime and derive the linearized hydrodynamic equations for the normal atomic gases with the NIST spin-orbit coupling for zero detuning $\delta=0$. We show that in the absence of the Galilean invariance, the hydrodynamics of the system crucially depends on the momentum susceptibilities, which can be modified by the spin-orbit coupling. We apply our general formalism to the ideal Fermi gases and reveal the effects of the spin-orbit coupling on the sound velocities and the dipole oscillation frequency of the gases.

\noindent \textbf{Results}

\noindent \textbf{Linearized Hydrodynamic Equations.} The normal atomic gases with the NIST spin-orbit coupling for zero detuning $\delta=0$ conserve the number of atoms $N$, the energy $E$, and the (pseudo-) momentum $\mathbf K$ [cf.~Eq.~(\ref{h0})]. The three conservation laws in their differential forms are
\begin{align}
&\frac{\partial \epsilon}{\partial t}+\nabla\cdot\mathbf j^{\epsilon}=0,\\
&\frac{\partial \mathbf g}{\partial t}+\nabla\cdot\mathbf \Pi=0,\\
&\frac{\partial  n}{\partial t}+\nabla\cdot\mathbf j^{ n}=0.
\end{align}
Here $\epsilon$, $\mathbf g$ and $ n$ are the local densities of energy, momentum and number of atoms respectively, and $\mathbf j^{\epsilon}$, $\mathbf \Pi$ and $\mathbf j^{ n}$ are the energy current, momentum current tensor and number current respectively. We assume that the inter-particle collision is so frequent that the system is in the hydrodynamic regime. The density of any physical quantity $O$ can be calculated by the local grand canonical ensemble with the distribution $\exp[-\beta(H-\mu N-\mathbf K\cdot \mathbf v)]$, where $\beta$, $\mu$ and $\mathbf v$ are the inverse of temperature $T$ (we take $k_{\rm B}=1$ throughout), chemical potential and velocity fields which are functions of position and time. The total Hamiltonian $H$ includes $H_0$ and interatomic interactions.
Note that due to the spin-orbit coupling, the spin degrees of freedom are not conserved. Spin dynamics in the presence of spin-orbit coupling has been studied in the context of both electron gases \cite{Sherman2010} and atomic Fermi gases \cite{Sherman2013}.

We focus on the dissipationless limit. The condition that the entropy change of the total system is zero determines the constitutive relations \cite{Martin, Chaikin}
\begin{align}
&\mathbf j^ n= n \mathbf v\\
&\Pi_{ab}=P\delta_{ab}+v_a g_b\\
&\mathbf j^\epsilon=(\epsilon+P)\mathbf v,
\end{align}
where $P$ is the pressure and the subscript stands for the component index. If the atoms are subject to an additional external potential $U_\text{ext}(\mathbf r)$, the momentum conservation equation becomes
\begin{align}
\frac{\partial g_b}{\partial t}+\sum_a\nabla_a(P\delta_{ab}+v_ag_b)+ n\nabla_b U_\text{ext}=0.
\end{align}

The absence of the Galilean invariance manifest through Eq.~(\ref{dispersion}) is a key feature of the atomic gas with spin-orbit coupling engineered by the Raman processes driven by external lasers. For normal atomic gases with the Galilean invariance, the momentum density $\mathbf g$ is always equal to $m n \mathbf v$ for arbitrary velocity field $\mathbf v$. However, this equality generally breaks down in the presence of spin-orbit coupling. The consequence of the lack of the Galilean invariance has been shown to affect the superfluidity and the bright solitons in Bose-Einstein condensates \cite{Wu1, Wu2}.
Nevertheless, if we are interested in small variations of $\beta$, $\mu$ and $\mathbf v$ away from the global equilibrium, we can keep to the first order of the variations and have 
\begin{align}
g_a(\beta,\mu, \mathbf v)=&\chi_{ab}v_b,\label{gv}\\
\epsilon(\beta,\mu, \mathbf v)=&\frac12 \sum_{a,b}\chi_{ab}v_av_b+\epsilon(\beta,\mu, \mathbf v=0),\label{ev}
\end{align}
where the momentum susceptibility is
\begin{align}
\chi_{ab}=\delta g_a/\delta v_b|_{\mathbf v=0}.
\end{align}
For simplicity, we have assumed that $\mathbf g(\beta,\mu,\mathbf v=0)=0$, i.e., when the gas is not moving, its momentum is zero. This assumption is true for the NIST scheme with zero detuning. Generalization of our results to cases without this assumption is straightforward.
Apparently $\chi_{ab}$ must be symmetric. We choose the coordinate system such that $\chi_{ab}=\tilde\chi_a\delta_{ab}$.
Using Eqs.~(\ref{gv}) and (\ref{ev}), we distill the linearized dissipationless hydrodynamic equations into
\begin{align}
\frac{\partial^2\delta n}{\partial t^2}=\sum_a\partial_a\left\{\frac{ n_0}{\tilde\chi_a}\left[\partial_a\left(\left(\frac{\partial P}{\partial  n}\right)_{\bar s}\delta n\right)+\delta n\partial_a U_{\rm ext} \right]\right\},\label{linhydro}
\end{align}
where $\delta n$ is the variation of number density $ n$ away from its global equilibrium value $ n_0$ and $({\partial P}/{\partial  n})_{\bar s}$ is taken at fixed entropy per particle. At $T=0$, Eq.~(\ref{linhydro}) further simplifies into
\begin{align}
\frac{\partial^2 \delta n}{\partial t^2}=\sum_a\partial_a\left[\frac{ n_0^2}{\tilde\chi_a}\partial_a\left(\frac{\partial\mu}{\partial n}\delta n\right)\right]\label{zerolinhydro}.
\end{align}
The phenomenological hydrodynamic equations (\ref{linhydro}) and (\ref{zerolinhydro}) shall be applicable to both normal Bose and Fermi gases with the spin-obit coupling.

\noindent \textbf{Momentum Susceptibility.}
The linearized hydrodynamic equations for atomic gases with spin-orbit coupling depend explicitly on
the momentum susceptibility $\chi_{ab}$ which can be calculated by the formula \cite{Chaikin}
\begin{align}
\chi_{ab}=\frac1V\sum_{f,i}\rho_i\frac{\langle i|K_b|f\rangle\langle f|K_a|i\rangle}{E_f-E_i},
\end{align}
where $E_i$ and $|i\rangle$ are the eigenvalues and eigenstates of the many-body Hamiltonian $H$, $\rho_i$ is the distribution function, and $V$ is the gas volume.

To manifest the effects of the spin-orbit coupling on $\chi_{ab}$, we assume that the inter-atomic interaction is weak enough that we can evaluate $\chi_{ab}$ using the ideal gas Hamiltonian, and find
\begin{align}
\chi_{ab}=\sum_{\alpha=\pm}\frac{ \partial}{\partial\mu}\int \frac{d^D\mathbf k}{(2\pi)^D}k_b k_a  f(\epsilon_{ \alpha,\mathbf k}),\label{chiideal}
\end{align}
where $f(\epsilon_{ \alpha,\mathbf k})$ is the Fermi (Bose) distribution function for fermionic (bosonic) atoms, $D$ is the dimension of the gas. The information of the spin-orbit coupling is encoded in the single atom dispersion $\epsilon_{\alpha,\mathbf k}$. For degenerate Fermi gases, $\chi_{ab}$ shall be dominated by the contribution from close to the Fermi surface.

We calculate $\chi_{xx}$ ($\tilde\chi_{x}$) by Eq.~(\ref{chiideal}) for $1$D, $2$D and $3$D Fermi gases with spin-orbit coupling generated in the $x$ direction by the NIST scheme with $\delta=0$. The (quasi-) $1$D or $2$D gases can be achieved when there is strong confinement in the $y$ direction or in both the $y$ and $z$ directions. For the $1$D Fermi gas of chemical potential $\mu$ at zero temperature, we find \newline
1) $\mu/\epsilon_\text{r}\geq1+\Omega/2\epsilon_\text{r}$,\\
\begin{align}
\frac{\chi_{xx}}{mn}=1+\frac{\mu/\epsilon_\text{r}+1
-\sqrt{(\mu/\epsilon_\text{r}-1)^{2}-(\Omega/2\epsilon_\text{r})^{2}}}{2[\mu/\epsilon_\text{r}+(\Omega/4\epsilon_\text{r})^{2}]}
,\label{chi1Da}
\end{align}
2) $1-\Omega/2\epsilon_\text{r}\leq\mu/\epsilon_\text{r}<1+\Omega/2\epsilon_\text{r}$,\\
\begin{align}
\frac{\chi_{xx}}{mn}=1+\frac{1}{\sqrt{\mu/\epsilon_\text{r}+(\Omega/4\epsilon_\text{r})^{2}}}
,\label{chi1Db}
\end{align}
3) $-(\Omega/4\epsilon_\text{r})^{2}\leq\mu/\epsilon_\text{r}<1-\Omega/2\epsilon_\text{r}$,\\
\begin{align}
\frac{\chi_{xx}}{mn}=1+\frac{\mu/\epsilon_\text{r}+1
+\sqrt{(\mu/\epsilon_\text{r}-1)^{2}-(\Omega/2\epsilon_\text{r})^{2}}}{2[\mu/\epsilon_\text{r}+(\Omega/4\epsilon_\text{r})^{2}]}
,\label{chi1Dc}
\end{align}
where the recoil energy is $\epsilon_{\rm r}=k_{\rm r}^2/2m$.

Figure~(\ref{chiomegaonezero}) shows at zero temperature how $\chi_{xx}$ for the $1$D Fermi gas changes with $\Omega$ for different atomic densities $n$. The plateaus appearing at small $\Omega$ are peculiar to this $1$D case and can be understood in the following way. When $\delta=0$ and $\Omega=0$, the two energy bands given by Eq.~(\ref{dispersion}) touch at energy $\epsilon_{\rm r}$ when $k_x=0$. The Fermi surface crosses the lower band at four points for $n/k_\text{r}<2/\pi(\approx0.64)$, and crosses both bands each at two points for $n/k_\text{r}>2/\pi$. When $\Omega$ is increased from zero, an energy gap $\sim\Omega$ opens at $k_x=0$. One can show analytically that for small densities $n$ before the point of the lower band at $k_x=0$ becomes lower than the Fermi surface, or for large densities $n$ before the upper band bottom at $k_x=0$ becomes higher than the Fermi surface, $\chi_{xx}/mn=1+(2k_\text{r}/\pi n)^2$. The further away the density $n$ is from $2/\pi$, the larger $\Omega$ the plateau persists to. Of course for our hydrodynamic approach to be applicable, $\Omega$ must be sufficiently larger than local equilibration rates. In the large $\Omega$ limit, the Fermi surface crosses the lower band at two points and $\chi_{xx}/mn\to(1-4\epsilon_\text{r}/\Omega)^{-1}$ [cf.~Eq.~(\ref{dispersion})].

%------------------------------------------------------------------------------

We also plot $\chi_{xx}$ versus atomic density $n$ for the $1$D Fermi gas with different $\Omega$ in Fig.~(\ref{chionezero}). When the density is high, the Fermi surface lies at high momenta where the spin-orbit coupling has negligible effects, and $\chi_{xx}$ approaches $m n$ as expected. When $\Omega/\epsilon_\text{r}=2$, $\epsilon_{-,k_x}$ has minima at two distinct $k_x$. For $\Omega/\epsilon_\text{r}=4$, $\epsilon_{-,k_x}$ has zero curvature at its single minimum. These features of $\epsilon_{-,k_x}$ give rise to the corresponding divergences of $\chi_{xx}/m n$ in the low density limit shown in Fig.~(\ref{chionezero}). When $\Omega/\epsilon_\text{r}=5$, the ratio $\chi_{xx}/m n$, though finite, is enhanced to be substantially larger than unity by the spin-orbit coupling. Similar behavior of $\chi_{xx}$ would be found in the $2$D and $3$D Fermi gases as well.

In Fig.~(\ref{chifinite}) is shown the finite temperature behavior of $\chi_{xx}$ for the Fermi gases with various Rabi frequency $\Omega$ in different dimensions. The spin-orbit coupling moves $\chi_{xx}/m n$ more away from unity at lower temperatures or in lower dimensions. When $T$ is finite, the low density limit corresponds to the chemical potential $\mu\to-\infty$. In this limit, to zero order, the distribution function $f$ in Eq.~(\ref{chiideal}) can be approximated by the Boltzmann distribution function; $\chi_{xx}/m n$ acquires the same value for fixed $\Omega$ and $T$ in different dimensions.

\noindent \textbf{Sound Velocities.}
In the case that there are no external potentials, i.e., $U_{\rm ext}=0$, we can read off the sound velocities along the principal axes from Eq.~(\ref{linhydro}) as
\begin{align}
c_a^2=\frac{ n_0}{\tilde\chi_a}\left(\frac{\partial P}{\partial  n}\right)_{\bar s},\label{ca}
\end{align}
where $\bar s$ is the entropy per particle. Besides the momentum susceptibility $\tilde\chi_a$, the sound velocities also depend on the equation of state via the adiabatic compressibility $\kappa_s=\left(\partial n/\partial P\right)_{\bar{s}}/n$. For convenience, we recast the sound velocities into
\begin{eqnarray}\label{CTD}
c_a^{2}=\frac{1}{\tilde\chi_{a}\kappa_{T}}\frac{C_{P}}{C_{V}},
\end{eqnarray}
by using the thermodynamic identity $\kappa_{s}/\kappa_{T}=C_{V}/C_{P}$, where $\kappa_{T}=\left(\partial n/\partial \mu\right)_{T,N}/n^{2}$ is the isothermal compressibility, $C_V=\left(\partial E/\partial T\right)_{V,N}$ and $C_P=\left(\partial E/\partial T\right)_{P,N}=C_V+VT\kappa_{T}\left(\partial P/\partial T\right)^{2}_{V,N}$ are the specific heats.

We calculate $c_x$ by Eq.~(\ref{CTD}) combined with our previous results for $\chi_{xx}$
for $1$D, $2$D and $3$D ideal Fermi gases with spin-orbit coupling generated in the $x$ direction by the NIST scheme with $\delta=0$. For convenience, the sound velocity is normalized by $c_0$ which is the corresponding sound velocity of the ideal Fermi gas without the spin-orbit coupling at the same density. Figure (\ref{sound1DTw4}) shows that at finite temperatures $c_x$ for the 1D case changes non-monotonically as the density of the Fermi gas varies. This non-monotonic behavior can be understood from the zero temperature limit. At zero temperature, according to the Gibbs-Duhem relation we have $dP/dn=nd\mu/dn$; from Eq.~(\ref{ca}) the sound velocities depend on the density of states at the energy scale of the chemical potential $\mu$. In 1D, when the chemical potential approaches the minimum of the NIST dispersion $\epsilon_{+,\mathbf k}$, the density of states is divergent. This divergence suppresses the sound velocity to zero as shown in Fig.~(\ref{sound1DTw4}) for $\Omega/\epsilon_\text{r}=4$, and gives rise to a discontinuity there. Therefore at finite temperatures, the thermal effects smear out the singular behavior of $c_x$ at $T=0$ and result in the non-monotonic one shown in Fig.~(\ref{sound1DTw4}).
When the dimension of the gas is raised up to 2D and 3D, the effects of the density of states on $c_x$ persist and cause a finite jump in 2D shown in Fig.~(\ref{sound2DTw4}) and a cusp in 3D as in Fig.~(\ref{sound3DTw4}) at zero temperature. Thus the finite temperature behavior of $c_x$ follows the general trend of its zero temperature one.

\noindent \textbf{Dipole Mode in Harmonic Traps.}
Collective modes of atomic gases confined in harmonic traps $U_{\rm ext}=m\omega_0^2 \mathbf{r}^2/2$ are important physical observables.
To investigate how the spin-orbit coupling affects the dipole mode frequency $\omega_\text{d}$ of the normal atomic gases with the NIST spin-orbit coupling, instead of solving Eq.~(\ref{linhydro}), we adopt a variational formalism which is equivalent to the linearized hydrodynamic equation (\ref{linhydro}) \cite{Zilsel,Taylor,Gao}. We start with the action $A=\int dt L$ and the Lagrangian is
\begin{align}
L=&\int d^D\mathbf r\left\{\frac12\sum_a\tilde\chi_a v_a^2-\varepsilon-n U_{\rm ext}+\phi\left[\frac{\partial n}{\partial t}+\nabla\cdot(n \mathbf v)\right]\right.\nonumber\\
&\left.+\eta\left[\frac{\partial s}{\partial t}+\nabla\cdot(s \mathbf v)\right]\right\},
\end{align}
where $s$ is the entropy density, and $\phi$ and $\eta$ are the Lagrangian factors introduced to enforce the number conversation
$\partial n/\partial t+\nabla\cdot(n \mathbf v)=0$ and the dissipationless condition $\partial s/\partial t+\nabla\cdot(s \mathbf v)=0$ \cite{Zilsel,Taylor}.

To estimate the frequency of the dipole mode oscillating in the principal axis $x$ direction, we assume the ansatz $n(\mathbf r,t)=n_0(\mathbf r-x_0(t)\hat x)$ and $s(\mathbf r,t)=s_0(\mathbf r-x_0(t)\hat x)$ with $n_0$ and $s_0$ the functions at equilibrium. This ansatz is motivated by the fact that the dipole mode is mainly the ``center of mass" motion of the gas. The number conservation gives $\mathbf v=\partial x_0(t)\hat x/\partial t$, which also maintains the dissipationless condition. After substituting the above ansatz into the Lagrangian, we have
\begin{align}
L=&L_0+\frac12\left[\int d^D\mathbf r\tilde\chi_{x}\right] v_x^2-\frac12\left[\int d^D\mathbf rn_0\right]m\omega_0^2 x_0^2,\label{l}
\end{align}
where $L_0$ is the Lagrangian at equilibrium. From Eq.~(\ref{l}), we obtain
\begin{align}
\frac{\partial^2 x_0}{\partial t^2}=-\frac{m\int d^D\mathbf r  n_0}{\int d^D\mathbf r\tilde \chi_{x}}\omega_0^2 x_0.
\end{align}
The dipole mode frequency is
\begin{align}
\frac{\omega_\text{d}}{\omega_0}=\left(\frac{m\int d^D\mathbf r  n_0}{\int d^D\mathbf r\tilde \chi_{x}}\right)^{1/2}. \label{od}
\end{align}
The independence of $\omega_\text{d}$ on the equation of state originates from the ansatz we use. It is worth mentioning that within the local density approximation, the result (\ref{od}) agrees with the one given by the sum rule approach \cite{Stringari}.
Figure (\ref{dipole}) plots the dipole mode frequency of 1D, 2D and 3D normal Fermi gases of $N_\text{tot}$ number of fermions with the spin-orbit coupling by the NIST scheme with $\delta=0$. Due to the enhancement of $\tilde\chi_x$ compared to $mn$ shown above, the dipole mode frequency $\omega_\text{d}/\omega_0$ is generally suppressed below unity.

%------------------------------------------------------------------------------

\noindent \textbf{Discussion}

\noindent The hydrodynamic equations (\ref{linhydro}) and (\ref{zerolinhydro}) are valid for normal atomic gases with any spin orbit coupling under the condition that the momentum $\mathbf K=0$ if the velocity $\mathbf v=0$. Apparently even within the NIST scheme, if $\delta\neq0$, this condition does not hold. Equation (\ref{gv}) shall include first order variation of $\mu$ and $\beta$, and Eq.~(\ref{ev}) shall change correspondingly; the generalization to Eqs.~(\ref{linhydro}) and (\ref{zerolinhydro}) shall be straightforward.
We have revealed the effects of spin-orbit coupling on the hydrodynamics of atomic gases by explicitly applying our general formalism to ideal Fermi gases. Interatomic interactions are expected to make quantitative changes to the results presented above. However, to take into account the interaction effects, one needs reliable determination of the susceptibilities $\chi_{ab}$ and the equation of state of the gases, which is beyond the scope of our work. The generalization of our hydrodynamic equations to the cases in which there is condensation requires correct treatment of the ``superfluid" part. Since the form of the ``superfluid" part depends on the exact structure of the order parameter, we leave the generalization to a future study.

Our calculation manifests the effects of the spin-orbit coupling on physical observables such as sound velocities and dipole mode frequency. 
Previously the sound velocity of a unitary Fermi gas has been measured by creating a local density variation through a thin slice of green laser and monitoring the propagation of the density variation afterwards\cite{Thomas}. Recent measurement of second sound velocity in the superfluid phase of unitary Fermi gases has also been achieved \cite{Grimm}.
The dipole oscillation of a Bose-Einstein condensate with spin-orbit coupling was excited by a sudden change of the Raman detuning and yielded clear violation of Kohn theorem \cite{Chendipole}. We expect that similar experimental techniques can be employed to confirm our theoretical predictions.

%------------------------------------------------------------------------------

\noindent \textbf{Acknowledgements} We thank Zeng-Qiang Yu and Hui Zhai for helpful discussions. This work is supported by Tsinghua University Initiative Scientific Research Program, NSFC under Grant No.~11474179 and No.~11204152.\\

\noindent \textbf{Author Contributions} Y.H. and Z.Y. wrote the main manuscript text and Y.H. prepared Figures 1-7.
\\

\noindent \textbf{Additional information}\\
\textbf{Competing financial interests:} The authors declare no competing financial interests.

%------------------------------------------------------------------------------
\newpage
\begin{center}
\begin{figure}
\centering
\includegraphics[width=\columnwidth]{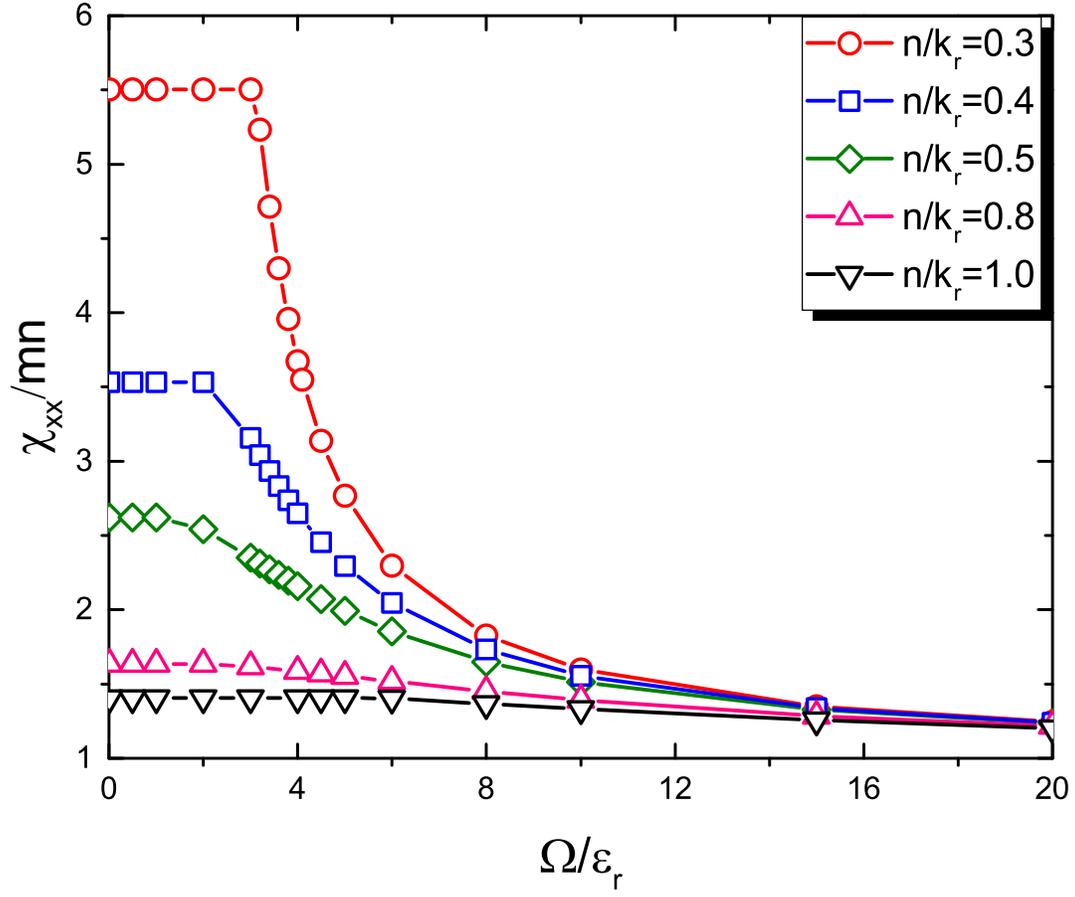}
\caption{ \label{chiomegaonezero}
\textbf{Momentum susceptibility $\chi_{xx}$ of the 1D ideal Fermi gas at zero temperature versus Rabi frequency $\Omega$}. Dependence of the momentum susceptibility $\chi_{xx}$ of the 1D ideal Fermi gas on $\Omega$ for different gas densities $n$.}
\end{figure}

\begin{figure}
\centering
\includegraphics[width=\columnwidth]{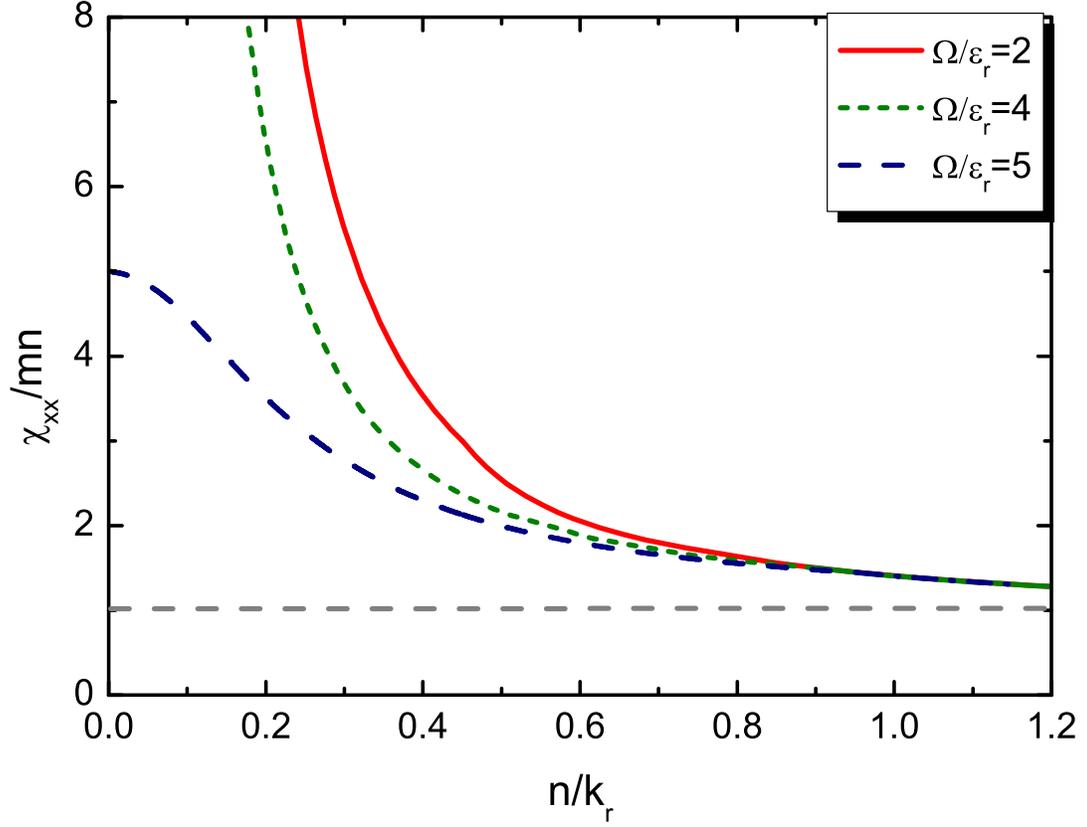}
\caption{ \label{chionezero}
\textbf{Momentum susceptibility $\chi_{xx}$ of the 1D ideal Fermi gas at zero temperature versus atomic density $n$}. Dependence of the momentum susceptibility $\chi_{xx}$ of the 1D ideal Fermi gas on the Rabi frequency $\Omega$ and the gas density $n$. The grey dash line is $\chi_{xx}=mn$ for cases in which the spin-orbit coupling has negligible effects.}
\end{figure}

\begin{figure}
\centering
\includegraphics[width=\columnwidth]{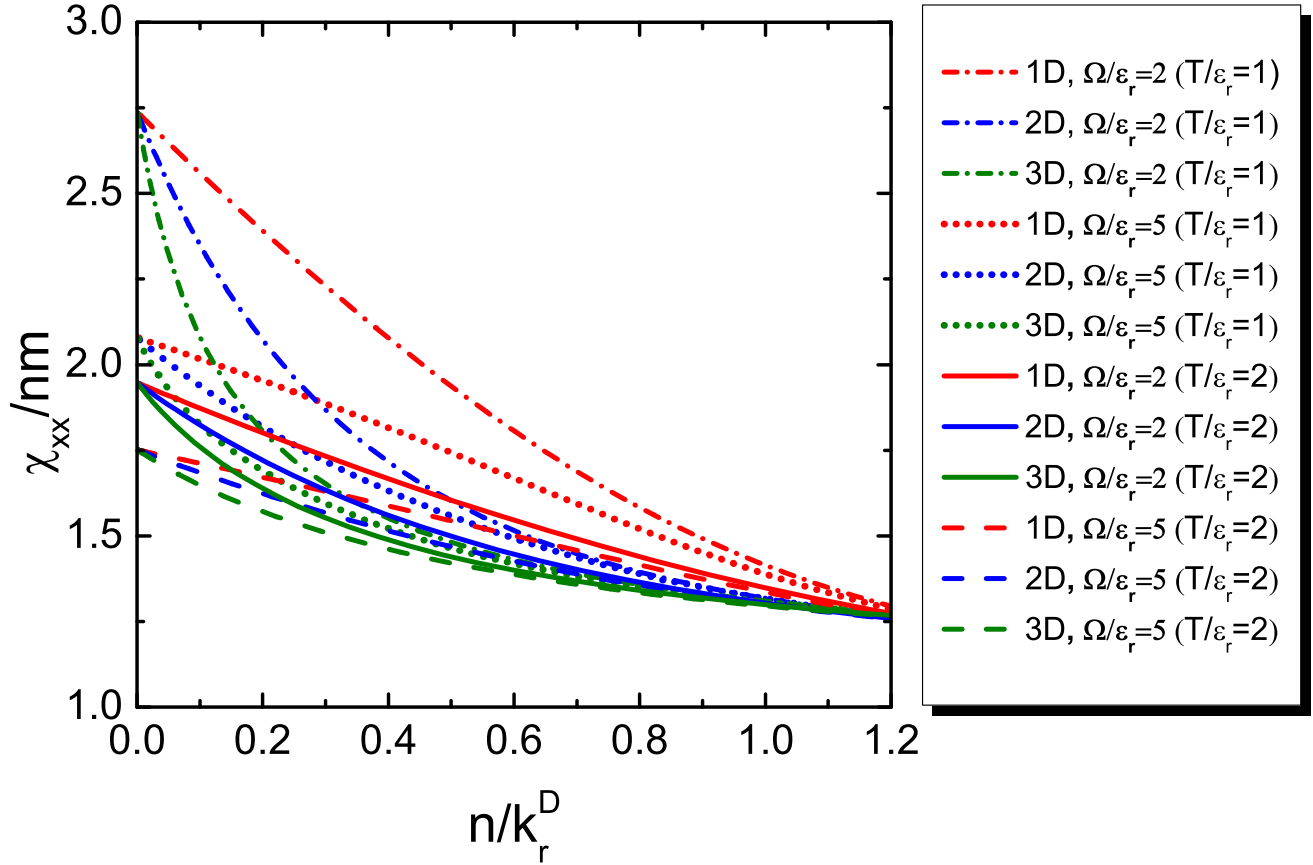}
 \caption{\label{chifinite}\textbf{Momentum susceptibility $\chi_{xx}$ of the 1D, 2D and 3D ideal Fermi gas at finite temperature versus atomic density $n$}. Dependence of $\chi_{xx}/mn$ on the temperature $T$ and the density $n$ for the 1D, 2D and 3D ideal Fermi gases.}
\end{figure}
\end{center}
\newpage

\begin{center}
\begin{figure}
\centering
\includegraphics[width=\columnwidth]{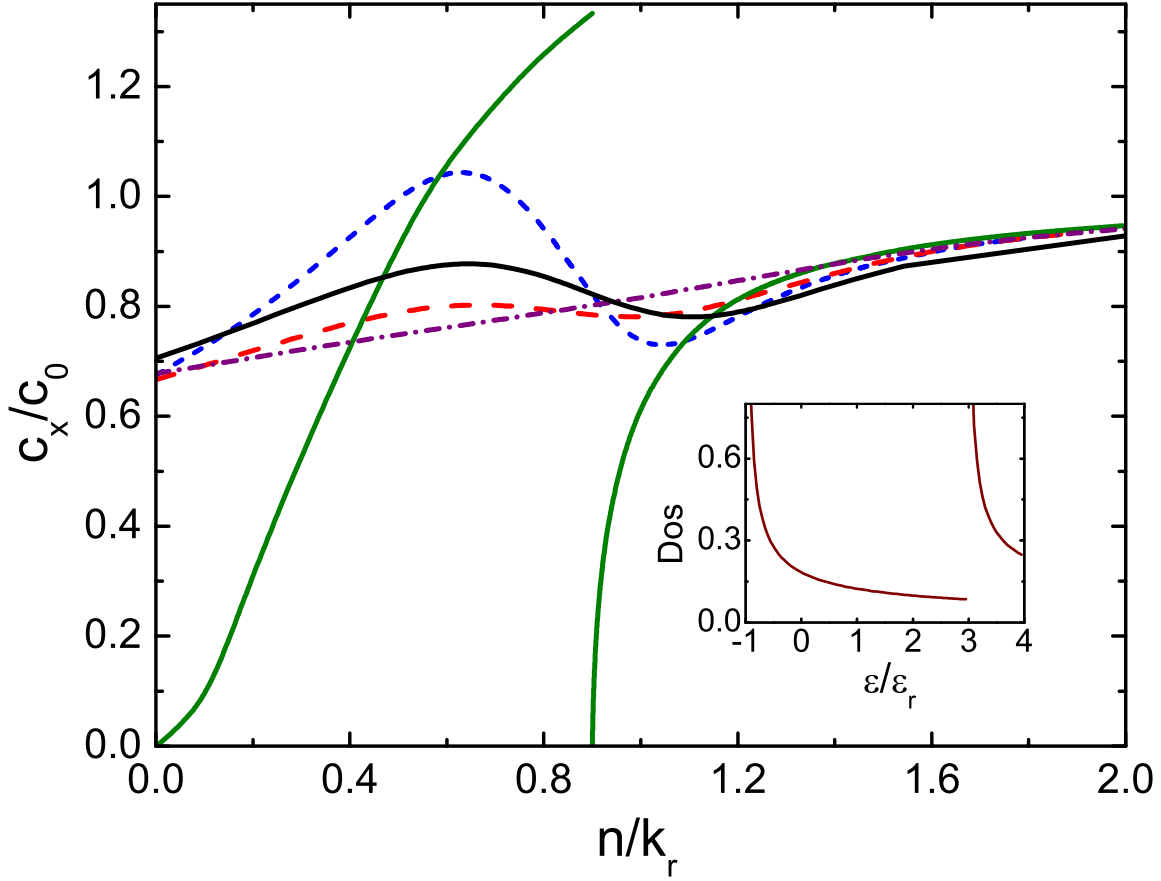}
 \caption{\textbf{Sound velocity $c_x$ of the 1D ideal Fermi gas versus atomic density $n$}. Dependence of sound velocity $c_x$ of the 1D ideal Fermi gas on the temperature $T$ and the density $n$. The sound velocity is normalized by $c_0$ the corresponding sound velocity of 1D ideal Fermi gas without the spin-orbit coupling at the same density. The green solid line is for $T=0$, the blue short--dash line for $T=0.5 \epsilon_\text{r}$, the red dash line for $T=1.0 \epsilon_\text{r}$, the purple short--dash--dot line for $T=2.0 \epsilon_\text{r}$ all with $\Omega=4\epsilon_\text{r}$. The black solid line is for $T=1.0 \epsilon_\text{r}$ with $\Omega=5\epsilon_\text{r}$. Inset is the density of states in unit of $V_{1D}k_\text{r}/\epsilon_\text{r}$ at $\Omega=4\epsilon_\text{r}$.}
  \label{sound1DTw4}
\end{figure}

\begin{figure}
\centering
\includegraphics[width=\columnwidth]{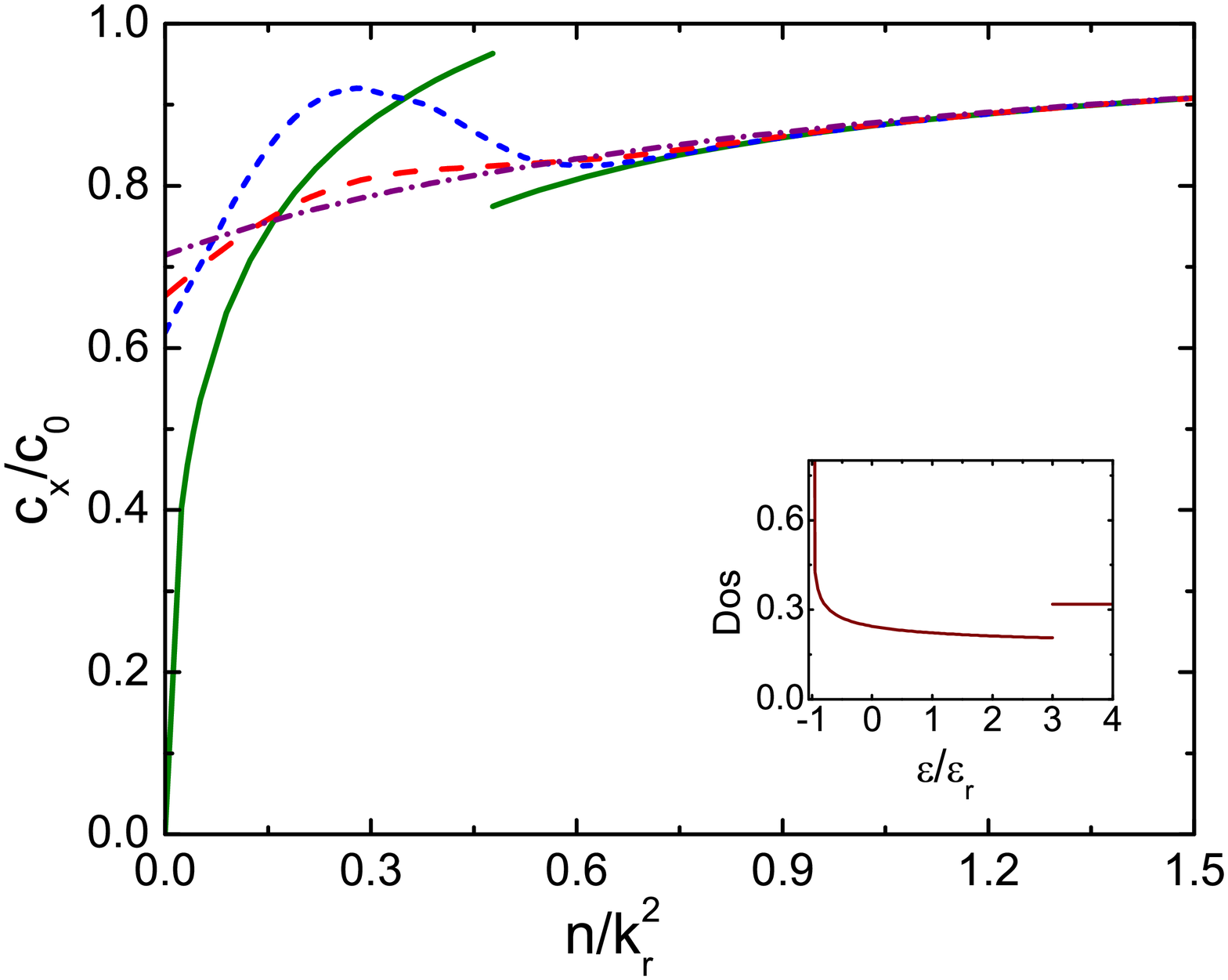}
 \caption{\textbf{Sound velocity $c_x$ of the 2D ideal Fermi gas versus atomic density $n$}. Dependence of sound velocity $c_x$ of the 2D ideal Fermi gas with $\Omega=4\epsilon_\text{r}$ on the temperature $T$ and the density $n$. Here $c_0$ is the finite temperature sound velocity of the 2D ideal Fermi gas without the spin-orbit coupling. See the caption of Fig.~\ref{sound1DTw4} for the temperatures for each curves. Inset is the density of states in unit of $mV_{2D}$ at $\Omega=4\epsilon_\text{r}$.}
  \label{sound2DTw4}
\end{figure}

\begin{figure}
\centering
\includegraphics[width=\columnwidth]{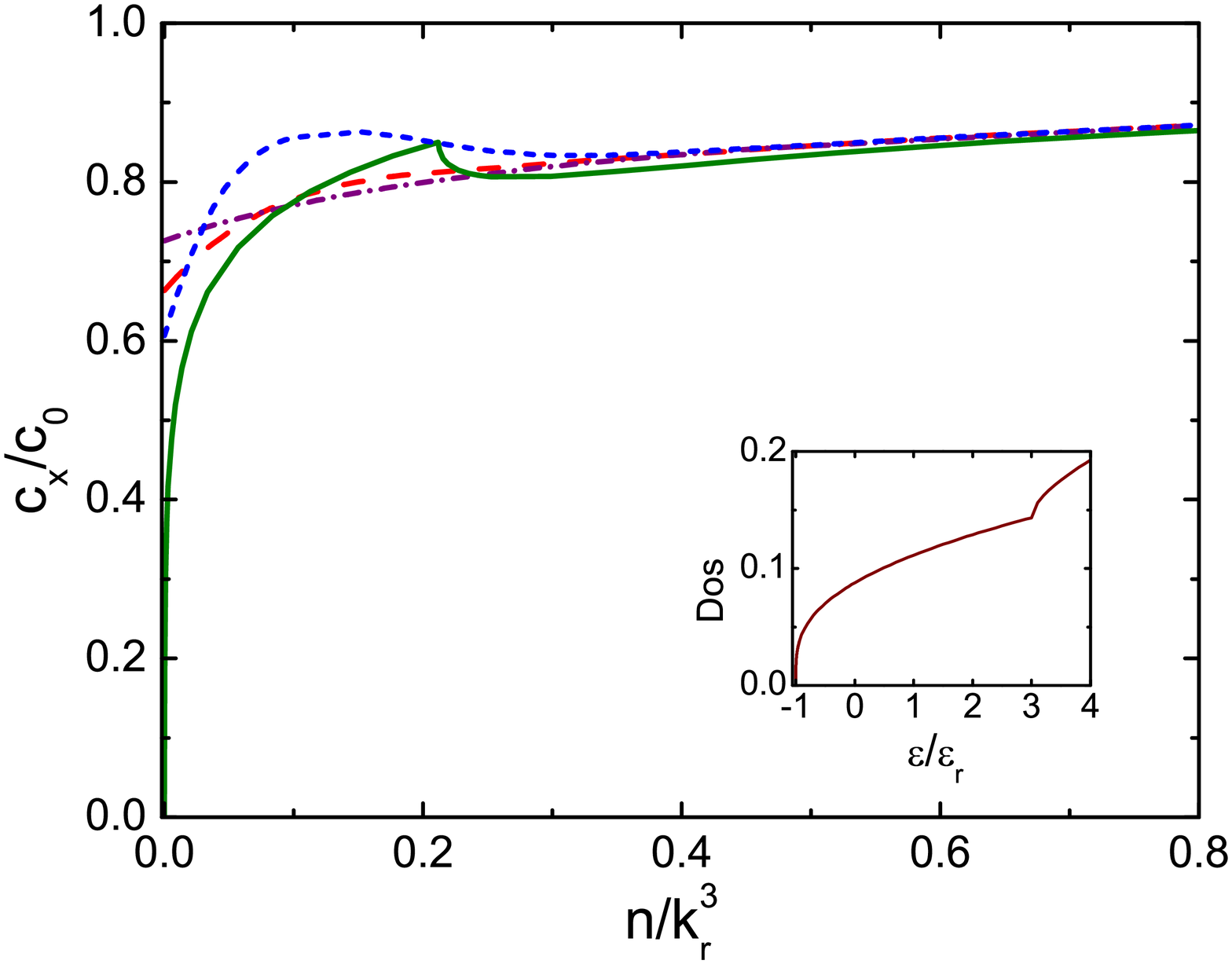}
 \caption{\textbf{Sound velocity $c_x$ of the 3D ideal Fermi gas versus atomic density $n$}. Dependence of sound velocity $c_x$ of the 3D ideal Fermi gas with $\Omega=4\epsilon_\text{r}$ on the temperature $T$ and the density $n$. Here $c_0$ is the finite temperature sound velocity of the 3D ideal Fermi gas without the spin-orbit coupling. See the caption of Fig.~\ref{sound1DTw4} for the temperatures for each curves. Inset is the density of states in unit of $mV_{3D}k_\text{r}$ at $\Omega=4\epsilon_\text{r}$.}
  \label{sound3DTw4}
\end{figure}
\end{center}
\newpage

\begin{figure}
\centering
\includegraphics[width=\columnwidth]{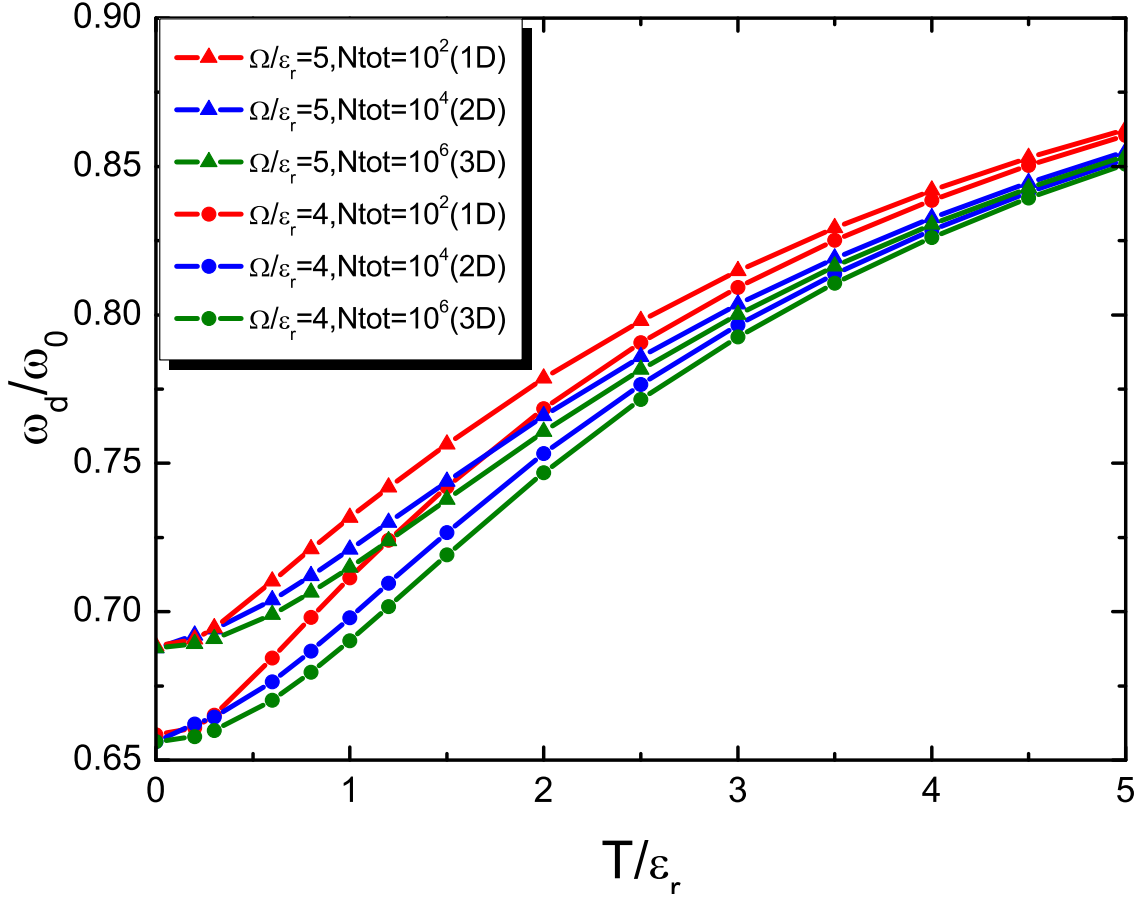}
 \caption{\textbf{Dipole mode frequency $\omega_{\rm d}$ versus temperature $T$}. Dependence of dipole mode frequency at finite temperature on the temperature and the Raman coupling for the 1D, 2D and 3D systems. We have taken $\omega_{0}=2\pi\times164$ Hz and $\epsilon_\text{r}=2\pi\times8340$ Hz \cite{Zhang} and the total number of particles $N_\text{tot}=10^2, 10^4, 10^6$, respectively.}
  \label{dipole}
\end{figure}

\end{document}